\documentclass[fleqn,twoside]{article}
\usepackage{espcrc2}

\usepackage{epsfig}
\usepackage[figuresright]{rotating}

\def\Journal#1#2#3#4{{#1} #2 (#4) #3}
\def\NPBPS{{Nucl. Phys.} B (Proc. Suppl.)}
\def\PLB{{Phys. Lett.}  B}
\def\JHEP{{Journ. of High Energy Physics}}
\def\IJMPC{{Int. Journ. of Mod. Physics}  C}

\def\PRD{{Phys. Rev.} D}

\newcommand{\AmS}{{\protect\the\textfont2
  A\kern-.1667em\lower.5ex\hbox{M}\kern-.125emS}}

\title{Unquenched Numerical Stochastic Perturbation Theory}

\author{F. Di Renzo\address{Dipartimento di Fisica, Universit\`a di Parma 
	and INFN, Gruppo Collegato di Parma, Italy},
        V. Miccio\addressmark $\,$
        and
        L. Scorzato\address{Theory Division, DESY, Hamburg, Germany}}
       
\begin{document}

\begin{abstract}
The inclusion of fermionic loops contribution in Numerical Stochastic 
Perturbation Theory (NSPT) has a nice feature: it does not cost so 
much (provided only that an FFT can be implemented in a fairly efficient 
way). Focusing on Lattice $SU(3)$, we report on the performance of the 
current implementation of the algorithm and the status of first 
computations undertaken. 
\vspace{1pc}
\end{abstract}

\maketitle

\section{Introduction}

At Lattice 2000 we discussed how to include fermionic loops contributions in 
Numerical Stochastic Perturbation Theory for Lattice $SU(3)$, an algorithm 
which we will refer to as UNSPT (Unquenched NSPT). Our main message here 
is that unquenching NSPT results in not such a heavy computational 
overhead, provided only that an $FFT$ can be implemented in a fairly efficient 
way. $FFT$ is the main ingredient in constructing the fermion propagator by 
inverting the Dirac kernel order by order. For a discussion of the foundations 
of UNSPT we refer the reader to \cite{NSPTlat00}. 

\begin{table*}[hbt]
\caption{Execution times of a fixed number of sweeps for quenched and unquenched NSPT 
(see main text for details).}
\label{table:1}
\newcommand{\m}{\hphantom{$-$}}
\newcommand{\cc}[1]{\multicolumn{1}{c}{#1}}
\renewcommand{\tabcolsep}{2pc} % enlarge column spacing
\renewcommand{\arraystretch}{1.2} % enlarge line spacing
\begin{tabular}{@{}lllll}
\hline
lattice size  - order - APEmille resource  & \cc{} & \cc{$N_f = 0$} & \cc{} & \cc{$N_f \neq 0$} \\
\hline
$8$ - $g^6$ - board                        & \m{} & \m29 & \m{} & \m50 \\
$8$ - $g^8$ - board                        & \m{} & \m51 & \m{} & \m85 \\
$8$ - $g^{10}$ - board                     & \m{} & \m82  & \m{} & \m134 \\
\hline
$16$ - $g^6$ - unit                         & \m{} & \m- & \m{} & \m212 \\
\hline
\end{tabular}\\[2pt]
\end{table*}

\section{Lattice SU(3) UNSPT on APEmille}

\begin{figure}[htb]
\vspace{9pt}
\begin{center}
\mbox{\epsfig{figure=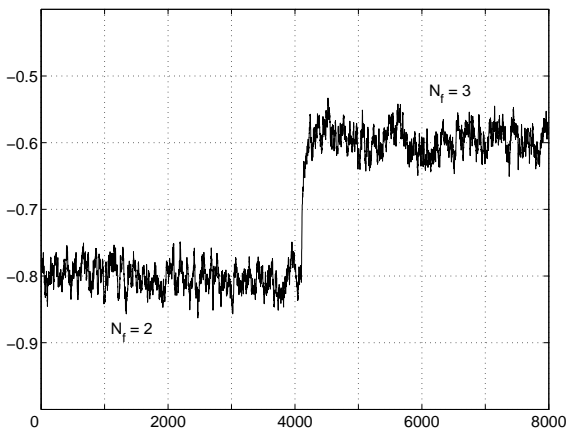,width=7.5cm}}
\caption{The effect of changing the number of flavours (on the fly).}
\label{fig:Nf_change}
\end{center}
\end{figure}

The need for an efficient $FFT$ is what forced us to wait for APEmille: our $FFT$ 
implementation 
mimic \cite{leleFFT}, which is based on a $1-dim \, FFT$ plus 
transpositions, an operation which asks for local addressing on a parallel 
architecture. 
UNSPT has been implemented both in single and in double precision, 
the former being remarkably robust for applications like Wilson loops. To estimate 
the computational overhead of unquenching NSPT one can inspect Table~\ref{table:1}. 
We report execution times of a fixed amount of sweeps both for quenched and 
unquenched NSPT. On both columns the growth of computational time is consistent with the 
the fact that every operation is performed order by order. On each row the growth 
due to unquenching is roughly consistent with a factor $5/3$. One then wants to understand 
the dependence on the volume, which is the critical one, the propagator being the 
inverse of a matrix: this is exactly the growth which has to be tamed by the $FFT$. 
One should compare execution times at a given order on $L=8$ and $L=16$ lattice sizes. 
Note that $L=8$ is simulated on an APEmille board ($8$ FPUs), while $L=16$ on an 
APEmille unit ($32$ FPUs). By taking this into account one easily understands that 
$FFT$ is doing its job: the simulation time goes as the volume also for UNSPT (a result 
which is trivial for quenched NSPT). Notice that at this level one has only compared 
crude execution times: a careful inspection of autocorrelations is anyway not going 
to jeopardize the picture. 
As for the dependence on $N_f$ (number of 
flavours), it is a parametric one: one plugs in various numbers and then proceed to fit 
the polynomial (in $N_f$) which is fixed by the order of the computation. It is then 
reassuring to find the quick response to a change in $N_f$ which one can inspect in 
Figure~\ref{fig:Nf_change} (which is the signal for second order of the plaquette at 
a given value of the hopping parameter $K$).

\section{Benchmark computation I: Wilson loops}

\begin{figure}[htb]
\vspace{9pt}
\begin{center}
\mbox{\epsfig{figure=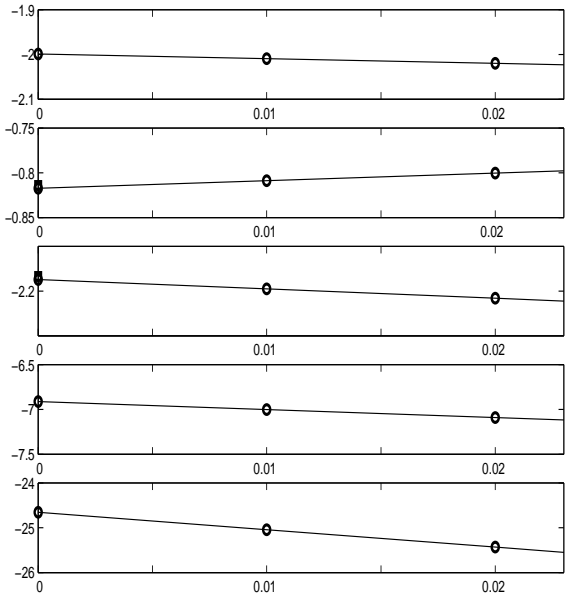,width=7.5cm,height=8cm}}
\caption{First five orders of the basic plaquette for $K=0.16$ and $N_f=2$ on a $L=8$ lattice. 
On $x$-axes different values of the time step $\tau$ used in integrating Langevin equation: 
the result has to be extrapolated to $\tau=0$. Analytic results for second and third 
order (first one is independent of sea fermions) in infinite volume are marked with a 
different symbol and they are hardly distinguishable.}
\label{fig:5ORDplaq}
\end{center}
\end{figure}
We now proceed to discuss some benchmark computations. A typical one is given by 
Wilson loops. In Figure~\ref{fig:5ORDplaq} one can inspect the first five 
orders \footnote{All our expansions are written in powers of $\beta^{-1}$.} 
of the basic plaquette at a given value of hopping parameter $K$, for which 
analytic results can be found in \cite{PIplaq}: going even higher in order would be 
trivial at this stage\footnote{Notice anyway that these computations are performed 
at a given value of hopping parameter $K$, but with no mass counterterm (see later).}. 
Apart for being an easy benchmark, we are interested in Wilson loops for two reasons. 
First of all we are completing the unquenched computation of the Lattice Heavy Quark 
Effective Theory Residual Mass (see \cite{ResMass} for the quenched result). On top of that 
we also keep an eye on the issue of whether one can explain in term of renormalons 
the growth of the coefficients of the plaquette. There is a debate going on about that 
(see \cite{RENorNOT}), the other group involved having also started to make use of 
NSPT. In the renormalon framework the effect of $N_f$ can be easily inferred from 
the $\beta$-function, eventually resulting in turning the series to oscillating signs.

\section{Benchmark computation II: the Wilson fermions Critical Mass} 

\begin{figure}[htb]
\vspace{9pt}
\begin{center}
\mbox{\epsfig{figure=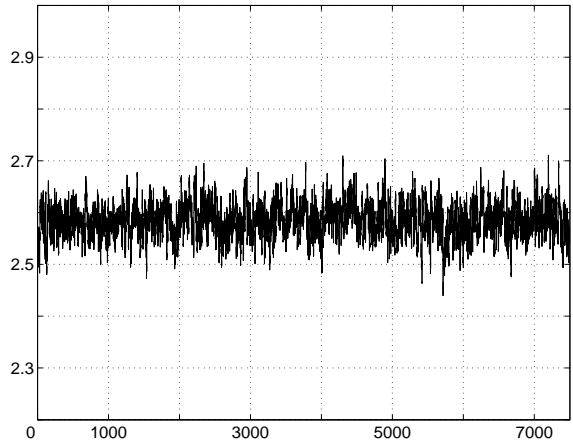,width=7.5cm,height=6cm}}
\caption{The signal for the first order of the Critical Mass on a $L=8$ lattice with a 
cutoff mass of $m'=0.01$, at a given value of the time step used in the integration 
of Langevin equation. Analytic result on the same volume and with the same cutoff 
is $2.557\ldots$}
\label{fig:MCg2}
\end{center}
\end{figure}
In Figure~\ref{fig:MCg2} we show the signal for one loop order of the Critical Mass 
for Wilson fermions (two loop results are available from \cite{ANAmc}). The computation 
is performed in the way which is the most standard in Perturbation Theory, {\em i.e.} by 
inspecting the pole in the propagator at zero momentum. This is already a 
tough computation. It is a zero mode, an $IR$ mass-cutoff is needed and the volume 
extrapolation is not trivial. On top of that one should keep in mind that also gauge 
fixing is requested. The coefficients which are known analytically can be reproduced. 
Still one would like to change strategy in order to go to higher orders (which is a 
prerequisite of all other high order computations). The reason is clear: we have actually 
been measuring the propagator $S$, while the physical information is actually coded in 
$\Gamma_2 = S^{-1}$ (one needs to invert the series and huge cancellations are on 
their way). Notice anyway that the fact that the Critical Mass is already known to two-loop 
makes many interesting computations already feasible.

\section{Conclusions}

Benchmark computations in UNSPT look promising, since the computational overhead 
of including fermionic loops contributions is not so huge. This is to be contrasted 
with the heavy computational effort requested for non perturbative unquenched lattice QCD. 
This in turn suggests the strategy of going back to perturbation theory for the (unquenched) 
computation of quantities like improvement coefficients and renormalisation constants. 
The Critical Mass being already known to two loops, many of these computations are 
already feasible at $\alpha^2$ order. \\
We have only discussed the implementation of the 
algorithm on the APEmille architecture. We can also rely on a $C^{++}$ implementation 
for PC's (clusters) which is now at the final stage of development.

\end{document}